# Short range order and topology of $Ge_xGa_xTe_{100-2x}$ glasses

I. Pethes[a,*], A. Piarristeguy[b], A. Pradel[b], S. Michalik[c], R. Nemausat[d],

J. Darpentigny[e], P. Jóvári[a]

[a]Wigner Research Centre for Physics, Institute for Solid State Physics, H-1525 Budapest, POB 49, Hungary

[b]Univ Montpellier, CNRS, ENSCM, Montpellier, France

[c]Diamond Light Source Harwell Science and Innovation Campus, Didcot, Oxfordshire, OX11 0DE, UK

[d]Deutsches Elektronen Synchrotron DESY, Notkestrasse 85, D-22607 Hamburg, Germany

[e]Laboratoire Léon Brillouin, CEA-Saclay 91191 Gif sur Yvette Cedex France

**Abstract**

Chemical short range order and topology of $Ge_xGa_xTe_{100-2x}$ glasses was investigated by neutron- and x-ray diffraction as well as Ge and Ga K-edge extended x-ray absorption fine structure (EXAFS) measurements. Large scale structural models were obtained by fitting experimental datasets simultaneously with the reverse Monte Carlo simulation technique. Models, relying only on experimental data and basic physical information without constraining the average coordination numbers, give 3.9 - 4.1 for the number of the atoms in the first coordination sphere of Ge atoms, while the average number of first neighbors of Ga atoms scatters around 3.8. The average coordination number of Te atoms is significantly higher than 2 for $x = 12.5$ and 14.3. It is found that the vast majority of $MTe_4$ (M=Ge or Ga) tetrahedra have at least one corner sharing $MTe_4$ neighbor.

**Introduction**

Due to their broad infrared transmission window glassy tellurides are extensively used in various fields of IR optics. The general strategy to find tellurides with excellent glass forming ability is to alloy the prototype Ge-Te system with a third component. Glasses with Ge-X-Te (X = Ga, As, Se, I, Ag, AgI) composition often possess a broad supercooled liquid region that makes it possible to shape bulk infrared lenses or draw fibers transmitting up to at least 18 μm [1 - 3]. It has been shown recently that in binary $Ge_xTe_{100-x}$ (14.5 ≤ $x$ ≤ 23.6) glasses the total average coordination numbers of Ge and Te atoms are – within the experimental uncertainty – 4 and 2, respectively [4]. It has also been revealed that Ge-Te glasses are chemically ordered: Ge-Te bonds are clearly preferred to Ge-Ge ones, even if the latter can be found in $Ge_{23.6}Te_{76.4}$. Alloying affects the structure of the host Ge-Te network in different ways. Se and I bind predominantly to Ge and do not change the average coordination numbers of Ge and Te [5, 6]. In Te-poor compositions As atoms bind to Ge, As and Te atoms but the average coordination numbers of Ge and Te atoms do not change here either [7]. On the other hand, in $GeTe_4$-AgI glasses the average coordination number of Te atoms is significantly higher than 2 even if only Ge/Te neighbors are taken into account [8]. Therefore, the topology of the host Ge-Te network changes significantly upon adding AgI.

[*] Corresponding author. E-mail address: pethes.ildiko@wigner.hu

The first experimental study of Ge-Ga-Te glasses combining diffraction techniques and EXAFS in the framework of reverse Monte Carlo (RMC) simulation technique [5] reported that the total coordination number of Te is 2.36 ± 0.15 while the average number of neighbors of Ga atoms is about 3 in $Ge_{11.1}Ga_{11.1}Te_{77.8}$ (the coordination number of Ge atoms was constrained to be 4). As the average coordination number of Ga was reported to be around 4 in several amorphous systems (e.g. $Ga_{50}Se_{50}$ [9], Ga-doped Ge:H [10], $CsCl$–$Ga_2S_3$ [11] and $CsCl$–$La_2S_3$–$Ga_2S_3$ [12]) the above study was followed by a further investigation of $Ge_{11.1}Ga_{11.1}Te_{77.8}$.

In the experimentally constrained density functional (DFT) study of Voleská et al [13] the starting configuration was obtained by fitting diffraction and EXAFS datasets simultaneously with RMC. This configuration was optimized by DFT and finally 'experimentally refined' again by RMC by using the DFT bond angle distributions as constraints. This configuration reasonably reproduced the experimental data and had a total energy only 33.8 meV/atom higher than that of the original DFT structure. The average coordination numbers of Ga, Ge and Te atoms were 4.08, 3.77 and 2.59, respectively.

While the coordination number of Ga is rather close to the values found in refs. [9 - 12] the average number of neighbors of Te atoms is significantly higher than the experimentally determined coordination number (2.36 ± 0.15). We note here that due to its high concentration in $Ge_{11.1}Ga_{11.1}Te_{77.8}$ the average coordination number of Te can be deduced from experimental data with a relatively low uncertainty.

The discrepancy of experimental (RMC) and DFT values is mainly due to the shallow minimum of the Te-Te partial pair correlation function in the DFT-generated model. More recent DFT studies emphasized the importance of the choice of exchange-correlation functionals and the proper treatment of van der Waals interactions [14 - 16]. It was demonstrated that by using Becke-Lee-Yang-Parr (BLYP) exchange-correlation functional and van der Waals forces in modelling amorphous tellurides some problems of earlier DFT simulations (e.g. high number of Ge atoms in octahedral environment in a covalent system, too high bond distances) can be avoided. The total coordination number of Te in $GeTe_4$ is also closer to 2 though the deviation from the experiment-based value is still significant (2.31 vs. 2.00 ± 0.1 in $Ge_{18.7}Te_{81.3}$ [4]).

In case of Ge-Ga-Te glasses the main difficulty of experimental structure determination is that Ga and Ge possess similar scattering properties both for X-rays and neutrons ($Z_{Ge}$=32, $Z_{Ga}$=31, $b_{Ge}$=8.185 fm, $b_{Ga}$=7.288 fm, where $Z$ is the atomic number and $b$ is the coherent neutron scattering length). Another problem is that the mean Ga-Te nearest neighbor distance is between the Ge-Te and Te-Te bond lengths [14], therefore Ga-Te peak parameters (especially the coordination number) are more sensitive to the 'cross talk' between overlapping peaks.

Even if the uncertainty of structural parameters is relatively large for a single composition, reliable information can be obtained from experimental data by studying a concentration series. For this reason, we investigated the structure of some $Ge_xGa_xTe_{100-2x}$ glasses by combining X-ray and neutron diffraction data with Ge- and Ga K-edge EXAFS measurements in the framework of the reverse Monte Carlo simulation technique. Short range order parameters of Ge-Ga-Te glasses are compared with those of amorphous Ge-Te, Ge-Ga-S and Ge-Ga-Se alloys as well as with models of Ge-Ga-Te glasses obtained by ab initio molecular dynamics. Though the experimental data contain limited information on medium range order the evolution of the pair correlation functions allows us to draw some conclusions about ordering of $GeTe_4$ and $GaTe_4$ tetrahedra as well.

**Experimental**

Four Ge-Ga-Te glasses of nominal compositions $Ge_{7.5}Ga_{7.5}Te_{85}$, $Ge_{10}Ga_{10}Te_{80}$, $Ge_{12.5}Ga_{12.5}Te_{75}$, and $Ge_{14.3}Ga_{14.3}Te_{71.4}$ were used for both neutron and X-ray experiments. Starting elements from high-purity germanium pellets (99.999%, Goodfellow), gallium ingots (99.9995%, Sigma-Aldrich), and tellurium ingots (99.9999%, Sigma-Aldrich) were first weighed in stoichiometric quantities (for a total batch of ~3 g) and introduced in a cylindrical silica tube (11 mm inner diameter, 1 mm wall thickness). The tube was subsequently evacuated under secondary vacuum ($10^{-5}$ mbar), sealed and heated up to 1220 K in a furnace with a heating rate of 10 K/h. The molten batch was held at this temperature for three days and finally quenched in a salt−ice-water after an annealing step of two days at 1073 K.

Neutron diffraction (ND) measurements were carried out at the 7C2 diffractometer of LLB (Saclay, France). Powdered samples were filled into vanadium sample holders of 6 mm diameter and 0.1 mm wall thickness. The wavelength of incident neutrons was 0.723 Å. The wavelength and detector position were determined by measuring a standard Ni powder sample. Raw data were corrected for background scattering and detector efficiency.

High energy X-ray diffraction (XRD) measurement was carried out at the Joint Engineering, Environmental and Processing (I12-JEEP) beamline at Diamond Light Source Ltd (UK). The size of the monochromatic beam was $0.3 \times 0.3$ mm$^2$. A CeO$_2$ reference sample (NIST Standard Reference Material 674b) was measured at different distances to determine the energy of the incident beam, the sample-to-detector distance, the position of the beam centre and the tilt of the detector. The wavelength of the incident beam and the sample-to detector distance were 0.1255 Å (98.768 eV) and 336 mm, respectively. Collected 2D diffraction data were integrated into reciprocal-space using the DAWN software [17]. X-ray structure factor, $S^X(Q)$, were extracted from integrated raw data using the PDFGetX2 software [18].

Ge and Ga K-edge EXAFS spectra were measured in fluorescence mode at beamline P65 of the Petra III source. Samples were finely ground, mixed with cellulose and pressed into tablets. Monochromatic radiation was obtained by a Si(111) double crystal monochromator. $\chi(k)$ curves were obtained using the Viper program [19]. Raw $\chi(k)$ signals were first forward Fourier-transformed using a Kaiser-Bessel window. The resulting $r$-space curves were back transformed using a rectangular window over 1.1-2.4 Å.

**Reverse Monte Carlo simulations**

The reverse Monte Carlo (RMC) method [20] is robust tool to obtain large three-dimensional structural models consistent with the fitted (experimental and/or theoretical) data sets. It can be used with any quantity that can be obtained from the atomic coordinates, such as total structure factors from ND or XRD experiments or EXAFS curves. A strength of the method is that the data sets can be fitted simultaneously. During the simulation particles are moved around randomly to minimize the differences between experimental and model curves. Finally particle configurations compatible with all fitted data sets (within the experimental error) are obtained. From these configurations short range order parameters (partial pair correlation functions, average coordination numbers etc.) can be calculated.

In the present study the RMC++ code [21] was used to produce structural models. The EXAFS backscattering coefficients were calculated by the FEFF8.4 program [22].

Investigated samples, their estimated densities and the fitted data sets are collected in Table 1. Densities were estimated using literature values of amorphous $Ge_xTe_{100-x}$ [23, 24] and $Ga_xGe_yTe_{100-x-y}$ glasses [25 - 29]. The simulation boxes contained 10000 atoms for the test runs and 40000 atoms for the final results. Initial configurations were obtained by placing the atoms randomly in the boxes and moving them around to satisfy the minimum interatomic distance (cutoff) requirements. Starting values of the cutoff distances were usually around 85-90% of the sum of the corresponding atomic radii ($r_{Ge} \approx 1.25$ Å, $r_{Ga} \approx 1.3$ Å, $r_{Te} \approx 1.4$ Å) [30], the final values are collected in Table 2. Ge-Te, Ga-Te and Te-Te bonds were allowed in all simulation runs. In the investigated samples the amount of Te atoms is more than twice of the amount of Ge + Ga atoms, thus besides the formation of $MTe_4$ (or $MTe_3$) units (M=Ge or Ga) Te-Te pairs must also be present, even if heteronuclear Ge-Te and Ga-Te bonds are preferred. The necessity of M-M type bonds was investigated by test runs for the highest M content $Ge_{14.3}Ga_{14.3}Te_{71.4}$ sample (see below). Allowing them had either no effect on fit quality or resulted in M-M type $g(r)$ functions with an artificially split first peak having maxima around 2.5 Å and 2.8 Å. From these test runs it was concluded that M-M type coordination numbers in glassy $Ge_{14.3}Ga_{14.3}Te_{71.4}$ are around or below the sensitivity of our method (about 0.3-0.4) and are presumably even smaller in the other compositions investigated.

In the final models Ge-Ge, Ge-Ga and Ga-Ga bonds were forbidden by using cutoff values higher than the expected bond lengths. In all simulation runs some low coordination numbers of the atoms (0 for Te, 0 and 1 for Ga, and 0, 1 and 2 for Ge) were eliminated. In the final models only the above coordination constraints were used. The quality of the fits of different models were compared via their 'goodness-of-fit' (*R*-factor) values:

$$R = \frac{\sqrt{\Sigma_i \left( S_{mod}(Q_i) - S_{exp}(Q_i) \right)^2}}{\sqrt{\Sigma_i S_{exp}^2(Q_i)}} \quad (1)$$

Here index *i* runs over the experimental points while 'mod' and 'exp' refer to model and experiment, respectively. A similar expression is valid for the EXAFS data.

**Results and discussion**

The experimental total structure factors ($S(Q)$) and filtered, $k^3$-weighted EXAFS curves ($k^3\chi(k)$) are shown in Figures 1-4. Also shown are the fits of the final models, in which only the Ge-Te, Ga-Te and Te-Te bonds were allowed and the average coordination numbers were not constrained. Partial pair correlation functions ($g_{ij}(r)$) obtained for the final model are shown in Fig. 5 while bond lengths and average coordination numbers are collected in Tables 3 and 4.

*Nearest neighbor distances*

The Ge-Te bond distances are around 2.60 Å, while the Ga-Te bond length is 2.62-2.63 Å for all glasses except $Ge_{10}Ga_{10}Te_{80}$ for which no EXAFS data were available. For this composition longer Ge-Te and Ga-Te distances are compensated by a shorter Te-Te bond length showing that diffraction data without EXAFS cannot completely separate Ge-Te, Ga-Te and Te-Te distances. For the other glasses $r_{GeTe}$ agrees well with previous results: in amorphous $Ge_xTe_{100-x}$ $r_{GeTe}$ = 2.59 Å was found by ND [31], 2.59-2.62 Å by EXAFS [32 - 36], 2.58-2.61 Å by combining diffraction, EXAFS and RMC techniques [4, 8, 24, 37, 38], 2.6 Å by anomalous X-ray scattering and RMC [39]. 2.60-2.64 Å Ge-Te bond distances were found in amorphous Ge-Sb-Te [40 - 42], 2.60 Å in Ge-As-Te [7] and 2.60-2.63 Å in Ge-Ga-Te glasses [5, 26].

Early density functional molecular dynamics simulations (DFT) resulted in longer Ge-Te bond lengths: 2.70 – 2.78 Å in $Ge_xTe_{100-x}$ and $Ge_xSb_yTe_{100-x-y}$ systems [23, 43 - 45]. The combination of DFT simulations with RMC refinement gave results closer to the experimental values: 2.65 Å in Ge-Ga-Te [13], 2.58 Å in $Ge_{15}Te_{85}$ [46]. Recent DFT simulations using the Becke-Lee-Yang-Parr (BLYP) exchange-correlation functional instead of Perdew-Burke-Ernzerhof (PBE), and especially with van der Waals dispersion forces included, have shown improved agreement with experimental values: 2.64-2.66 Å with PBE in $Ge_xTe_{100-x}$ [15, 47 - 49], 2.59-2.62 Å with BLYP in $Ge_xTe_{100-x}$ [15], 2.63 Å in $Ge_2Sb_2Te_5$ [50] and 2.59 Å in $Ge_{15}Ga_{10}Te_{75}$ [14].

The $r_{GaTe}$ value (2.62-2.63 Å) is the same as it was found experimentally in Ge-Ga-Te glasses by EXAFS [26] and by combining experimental (diffraction, EXAFS) data with simulation (RMC and density functional) [13]. Somewhat longer bond length was found in Ref. [14] (2.67 Å) with first principles molecular dynamics simulation (FPMD).

The Te-Te bond length is around 2.77 Å, except again for the $Ge_{10}Ga_{10}Te_{80}$ sample. Te-Te distances reported in the literature have a broad distribution: in amorphous $Ge_xTe_{100-x}$ 2.76 Å was measured by ND [31], 2.77-2.82 Å by EXAFS [33, 34], 2.70-2.79 Å by combination of diffraction, EXAFS and RMC [4, 5, 7, 8, 24, 37, 38], 2.73 Å by anomalous X-ray scattering and RMC [39]. Te-Te distances around 2.79-2.80 Å were reported in Ge-Ga-Te [5], 2.77 Å in Ge-As-Te [7], 2.77-2.79 Å in Ge-Te-Ag-I glasses [8] by combination of diffraction, EXAFS and RMC.

First principles molecular dynamics simulations result in longer Te-Te bond lengths: 2.87 Å [43], 2.85 Å [45], 2.90 Å [47, 48], 2.89 Å [49]. Shorter distances were obtained by DFT simulations with RMC refinement: 2.83 Å [13], 2.74 Å [46]. Treatment of van der Waals forces (VdW) seems to be important here as well: 2.83 -2.84 Å was obtained with and 2.88-2.89 Å without VdW in GeTe$_4$ [15] and 2.81 Å in Ge$_{15}$Ga$_{10}$Te$_{75}$ with BLYP+VdW [14].

*Average coordination numbers*

The average number of neighbours of Ge is around 4, as it is expected (and was reported earlier by experiments and simulations as well [4, 5, 7, 8, 13-15, 23, 26, 31, 33, 37, 39, 41-43, 45-51]). The average coordination number of Ga atoms is also close to 4, as was found in several Ge-Ga-Ch (Ch=S, Se, Te) glasses: experimentally in Ge-Ga-S glasses [52 - 59], in Ge-Ga-Se glasses [54, 60 - 65] and in amorphous Ge-Ga-Te [13, 26]. Recent FPMD simulations on 80GeSe$_2$-20Ga$_2$Se$_3$ glass [66], Ge-Ga-Te liquids [16] and Ge$_{15}$Ga$_{10}$Te$_{75}$ glass [14] also reported Ga coordination numbers around 4.

Test simulation runs were made in which coordination constraints were used to force Ge and Ga atoms to have exactly 4 neighbors (about 95% of the atoms satisfied this requirement). It was found that these coordination constraints have no effect on the quality of fits.

The total coordination number of Te increases with increasing Ge/Ga content (see Table 4). It is around 2 for the Ge$_{7.5}$Ga$_{7.5}$Te$_{85}$ glass and significantly higher than 2 for Ge$_{12.5}$Ga$_{12.5}$Te$_{75}$ and Ge$_{14.3}$Ga$_{14.3}$Te$_{71.4}$ (around 2.19 and 2.35, respectively). Test runs were carried out in which the Ge-Te and Ga-Te coordination numbers were constrained to remain 4 and the Te-Te coordination number was forced to decrease so that the total coordination number of Te be equal to 2. The *R*-factors of the fits for these models were significantly higher (mostly the *R*-factors of the ND and XRD data sets, with 20-50%), especially for Ge$_{12.5}$Ga$_{12.5}$Te$_{75}$ and Ge$_{14.3}$Ga$_{14.3}$Te$_{71.4}$. Besides the deterioration of the fits, the resulting Te-Te partial pair correlation functions exhibit artificially sharp peaks at around 3.1 Å, next to the upper limit of the coordination constraints.

Te coordination number around 2 was found in Ge$_x$Te$_{100-x}$ glasses by ND [31] and by EXAFS [34]. $N_{Te}$ = 2 was obtained by combination of diffraction and EXAFS experiments with RMC simulations in amorphous Ge$_x$Te$_{100-x}$ [4, 5, 7, 8, 24], in Ge-As-Te glasses [7] and in amorphous Ge-Sb-Te [41, 42].

FPMD simulations often result in Te coordination number higher than 2 (e.g. in Ge$_x$Te$_{100-x}$ [15, 23, 43, 47, 48, 49], in Ge-Sb-Te [43, 45, 51]). The actual value highly depends on the choice of the exchange correlation functional as well as on the usage of VdW forces (see e.g. Ref. [15]: 3.78 PBE, 3.51 PBE+VdW, 2.57 BLYP, 2.31 BLYP+VdW). Recent FPMD simulations proposed Te coordination number 2.9-4.8 in Ge-Ga-Te liquids [16], and 2.19 in Ge-Ga-Te glass [14]. The first minimum of $g_{TeTe}(r)$ obtained by these simulations is rather shallow thus the second coordination sphere may also contribute to the coordination number of Te.

As in binary Ge-Te glasses the coordination number of Te is around 2 [4] it is reasonable to assume that the increase observed in case of Ge-Ga-Te glasses is due to the presence of Ga atoms. It is to be noted that chemical ordering is also different in Ge-Te and Ge-Ga-Te glasses. While Ge-Ge bonds can be observed in melt quenched $Ge_{23.6}Te_{76.4}$ [4] no M-M type bonds were found in Ge-Ga-Te glasses. In principle, the investigated Ge-Te and Ge-Ga-Te glasses are all Te-rich therefore M-M bonding could be avoided even if all Te atoms remain twofold coordinated. Still, M-M bonds exist in $Ge_{23.6}Te_{76.4}$ and the coordination number of Te is higher than 2 in Ge-Ga-Te glasses. These two observations share a common consequence: they increase the Te-Te coordination number. Further experimental and theoretical studies are needed to see whether this is just a coincidence or a certain number of Te-Te bonds is favoured by the glassy state due to energetic or kinetic reasons. It is to be noted that the coordination number of S and Se is also higher than 2 in Ge-Ga-Se [64,65] and Ge-Ga-S glasses [57].

*Second neighbors*

As the mean M-Te distance is about 2.60 – 2.62 Å and the M-Te-M bond angles are centered around 90-95° [13, 46] it is reasonable to assume that the M-M peak at about 3.8 Å corresponds to the distance of two M atoms having a common Te neighbor. M atoms have predominantly 4 Te neighbors so it also means the existence of corner sharing (CS) $MTe_4$ tetrahedra. The presence of these motifs was confirmed by the analysis of configurations in which Ge and Ga atoms were constrained to have 4 Te neighbors. Edge sharing tetrahedra ($MTe_4$ units with two common Te atoms) were also formed during the simulation. Their existence can neither be confirmed nor excluded on the base of our experimental information. Images of such models of $Ge_{7.5}Ga_{7.5}Te_{85}$ and $Ge_{14.3}Ga_{14.3}Te_{71.4}$ glasses are shown in Figs. 6 and 7 while the statistics of CS and ES units is summarized in Table 5.

The decomposition of the first peak of the M-M partials to contributions of CS and ES tetrahedra and topologically distant pairs is presented for the Ge-Ge pairs of the $Ge_{14.3}Ga_{14.3}Te_{71.4}$ sample in Fig. 8 (Ga-Ga and Ge-Ga pairs show a similar behavior therefore only the Ge-Ga case is shown).

It was found that already for the $Ge_{7.5}Ga_{7.5}Te_{85}$ sample about 84% of $GeTe_4$ and $GaTe_4$ tetrahedra have at least one corner sharing $MTe_4$ neighbor (see Table 5). This value seems to be rather high in view of the low Ge/Ga-content of this glass. The average number of Ge and Ga atoms around Te is less than 1 meaning that formation of CS or ES pairs could be avoided, in principle. With increasing M content the number of the CS or ES tetrahedra also increases. For the $Ge_{14.3}Ga_{14.3}Te_{71.4}$ glass 99.7% of the $MTe_4$ tetrahedra have at least one CS or ES pair. The number of M atoms participating in ES units is around 12% for $Ge_{7.5}Ga_{7.5}Te_{85}$ and 33% for $Ge_{14.3}Ga_{14.3}Te_{71.4}$. We note here that the presence of $MTe_4$ units is the consequence of using coordination constraints (unconstrained simulations give configurations with more disordered local environments). On the other hand, the connectivity of those units stems from the position of the M-M peak (~3.8 Å), which clearly follows from the experimental data (see below).

*Prepeak in the ND total structure factor*

The neutron diffraction structure factor of $Ge_xGa_xTe_{100-2x}$ glasses has a first sharp diffraction peak (FSDP) or prepeak at $q_{max} \approx 1$ Å$^{-1}$. (A less pronounced peak can be observed in X-ray diffraction structure factor as well.) Peak positions and heights are given in Table 6. The height of the prepeak is defined as $S(q_{max}) - S(q_{min})$ where $q_{min}$ is the first minimum after the prepeak. For $Ge_{7.5}Ga_{7.5}Te_{85}$ there is only a shoulder therefore we used the $q_{min}$ value of $Ge_{10}Ga_{10}Te_{80}$. It can be observed that in case of $Ge_xGa_xTe_{100-x}$ glasses the height of the prepeak increases with increasing Ge/Ga content.

The connection of medium range order and prepeak intensity is confirmed by comparing the models of $Ge_{14.3}Ga_{14.3}Te_{71.4}$ obtained with and without fitting neutron diffraction data. Some results of these runs are shown in Fig. 9. It can be observed that models obtained without fitting neutron diffraction data fail to reproduce the prepeak of neutron diffraction structure factors. The other effect of omitting neutron diffraction data from the models is the rather flat first peak of the Ge-Ge partial pair correlation function. A similar behavior can be observed in $Ge_{18.7}Te_{81.3}$ [4], as displayed in Fig. 9. The first peaks of Ge-Ga and Ga-Ga partial pair correlation functions are affected in the same way (not shown). These observations strongly suggest that prepeak in $Ge_xGa_xTe_{100-2x}$ and Ge-Te glasses is connected to the well-defined M-M correlations (M = Ge, Ga) manifested in sharp real space M-M peaks. The latter, on the other hand, are the consequence of corner or edge sharing $MTe_4$ tetrahedra.

**Conclusions**

Short range order and topology of $Ge_xGa_xTe_{100-2x}$ ($x$ = 7.5, 10, 12.5, 14.3) glasses was investigated by diffraction techniques and EXAFS. Structural models were obtained by fitting experimental datasets simultaneously in the framework of the reverse Monte Carlo simulation technique. It was shown that Ga and Ge atoms are mostly fourfold coordinated while $N_{Te}$, the average coordination number of Te increases with Ge/Ga content ($N_{Te}$ = 2.35 ± 0.1 for $x$ = 14.3). The position of the first peak of M-M partial pair correlation functions (M=Ge, Ga) at 3.8 Å evidences the presence of corner or edge sharing tetrahedral units. It was found that already at $x$ = 7.5 the majority of Ge/Ga atoms are linked to other Ge/Ga atoms via one or two common Te neighbors. The prepeak of the neutron diffraction structure factors of $Ge_xGa_xTe_{100-2x}$ glasses is due to enhanced M-M correlations.

**Acknowledgment**

I. P. and P. J. were supported by NKFIH (National Research, Development and Innovation Office) Grant No. KH 130425. The neutron diffraction experiment was carried out at the ORPHÉE reactor, Laboratoire Léon Brillouin, CEA-Saclay, France. EXAFS measurements were carried out at PETRA III at DESY, a member of the Helmholtz Association (HGF). We thank Diamond Light Source Ltd. for access to beamline I12-JEEP (cm19662-1) that contributed to the results presented here.

**References**

[1] B. Bureau, S. Danto, H.L. Ma, C. Boussard-Plédel, X.H. Zhang, J. Lucas, Tellurium based glasses: A ruthless glass to crystal competition, Solid State Sci. 10 (2008) 427–433. doi:10.1016/j.solidstatesciences.2007.12.017.

[2] C. Conseil, J.-C. Bastien, C. Boussard-Plédel, X.-H. Zhang, P. Lucas, S. Dai, et al., Te-based chalcohalide glasses for far-infrared optical fiber, Opt. Mater. Express 2 (2012) 1470. doi:10.1364/OME.2.001470.

[3] Z. Yang, P. Lucas, Tellurium-based far-infrared transmitting glasses, J. Am. Ceram. Soc. 92 (2009) 2920–2923. doi:10.1111/j.1551-2916.2009.03323.x.

[4] P. Jóvári, A. Piarristeguy, A. Pradel, I. Pethes, I. Kaban, S. Michalik, et al., Local order in binary Ge-Te glasses – An experimental study, J. Alloys Compd. 771 (2019) 268–273. doi:10.1016/j.jallcom.2018.08.323.

[5] P. Jóvári, I. Kaban, B. Bureau, A. Wilhelm, P. Lucas, B. Beuneu, et al., Structure of Te-rich Te-Ge-X (X = I, Se, Ga) glasses, J. Phys.: Condens. Matter 22 (2010) 404207. doi:10.1088/0953-8984/22/40/404207.

[6] L. Rátkai, C. Conseil, V. Nazabal, B. Bureau, I. Kaban, J. Bednarcik, et al., Microscopic origin of demixing in $Ge_{20}Se_xTe_{80-x}$ alloys, J. Alloys Compd. 509 (2011) 5190–5194. doi:10.1016/j.jallcom.2011.02.032.

[7] P. Jóvári, P. Lucas, Z. Yang, B. Bureau, I. Kaban, B. Beuneu, et al., Short-Range Order in Ge-As-Te Glasses, J. Am. Ceram. Soc. 97 (2014) 1625–1632. doi:10.1111/jace.12823.

[8] P. Jóvári, S. Cui, V. Nazabal, I. Kaban, B. Beuneu, M. Dussauze, et al., Network Rearrangement in AgI-Doped $GeTe_4$ Glasses, J. Am. Ceram. Soc. 98 (2015) 1034–1039. doi:10.1111/jace.13369.

[9] K.D. Machado, P. Jóvári, J.C. de Lima, C.E.M. Campos, T.A. Grandi, Extended x-ray absorption fine structure, x-ray diffraction and reverse Monte Carlo studies of an amorphous $Ga_{50}Se_{50}$ alloy produced by mechanical alloying, J. Phys.: Condens. Matter 16 (2004) 581–590. doi:10.1088/0953-8984/16/4/007.

[10] I. Chambouleyron, D. Comedi, G. Dalba, P. Fornasini, R. Grisenti, F. Rocca, Internal stress-induced changes of impurity coordination and doping mechanisms in a-Ge:H doped with column III metals, Solid State Commun. 115 (2000) 89–93. doi:10.1016/S0038-1098(00)00130-7.

[11] F. Hindle, M. Miloshova, E. Bychkov, C.J. Benmore, A.C. Hannon, Structural analysis of $xCsCl(1-x)Ga_2S_3$ glasses, J. Non-Cryst. Solids 354 (2008) 134–137. doi:10.1016/j.jnoncrysol.2007.07.031.

[12] A.Y. Ramos, N. Watanabe, O.L. Alves, L.C. Barbosa, Structural characterisation of CsCl incorporation in $Ga_2S_3$–$La_2S_3$ glasses, J. Non-Cryst. Solids 304 (2002) 182–187. doi:10.1016/S0022-3093(02)01020-7.

[13] I. Voleská, J. Akola, P. Jóvári, J. Gutwirth, T. Wágner, T. Vasileiadis, et al., Structure, electronic, and vibrational properties of glassy $Ga_{11}Ge_{11}Te_{78}$: Experimentally constrained density functional study, Phys. Rev. B 86 (2012) 094108. doi:10.1103/PhysRevB.86.094108.

[14] Z. Chaker, G. Ori, M. Boero, C. Massobrio, E. Furet, A. Bouzid, First-principles study of the atomic structure of glassy $Ga_{10}Ge_{15}Te_{75}$, J. Non-Cryst. Solids 498 (2018) 338–344. doi:10.1016/j.jnoncrysol.2018.03.039.

[15] A. Bouzid, C. Massobrio, M. Boero, G. Ori, K. Sykina, E. Furet, Role of the van der Waals interactions and impact of the exchange-correlation functional in determining the structure of glassy $GeTe_4$, Phys. Rev. B 92 (2015) 134208. doi:10.1103/PhysRevB.92.134208.

[16] H. Flores-Ruiz, M. Micoulaut, A. Piarristeguy, M.-V. Coulet, M. Johnson, G.J. Cuello, et al., Structural, vibrational, and dynamic properties of Ge-Ga-Te liquids with increasing connectivity: A combined neutron scattering and molecular dynamics study, Phys. Rev. B 97 (2018) 214207. doi:10.1103/PhysRevB.97.214207.

[17] J. Filik, A.W. Ashton, P.C.Y. Chang, P.A. Chater, S.J. Day, M. Drakopoulos, et al., Processing two-dimensional X-ray diffraction and small-angle scattering data in DAWN 2, J. Appl. Crystallogr. 50 (2017) 959–966. doi:10.1107/S1600576717004708.

[18] X. Qiu, J.W. Thompson, S.J.L. Billinge, PDFgetX2 : a GUI-driven program to obtain the pair distribution function from X-ray powder diffraction data, J. Appl. Crystallogr. 37 (2004) 678–678. doi:10.1107/S0021889804011744.

[19] K. V Klementev, Extraction of the fine structure from x-ray absorption spectra, J. Phys. D: Appl. Phys. 34 (2001) 209–217. doi:10.1088/0022-3727/34/2/309.

[20] R.L. McGreevy, L. Pusztai, Reverse Monte Carlo simulation: A new technique for the determination of disordered structures, Mol. Simul. 1 (1988) 359–367. doi:10.1080/08927028808080958.

[21] O. Gereben, P. Jóvári, L. Temleitner, L. Pusztai, A new version of the RMC++ Reverse Monte Carlo programme, aimed at investigating the structure of covalent glasses, J. Optoelectron. Adv. Mater. 9 (2007) 3021–3027.

[22] A.L. Ankudinov, B. Ravel, J.J. Rehr, S.D. Conradson, Real-space multiple-scattering calculation and interpretation of x-ray-absorption near-edge structure, Phys. Rev. B 58 (1998) 7565–7576. doi:10.1103/PhysRevB.58.7565.

[23] J. Akola, R.O. Jones, Binary Alloys of Ge and Te: Order, Voids, and the Eutectic Composition, Phys. Rev. Lett. 100 (2008) 205502. doi:10.1103/PhysRevLett.100.205502.

[24] P. Jóvári, A.A. Piarristeguy, R. Escalier, I. Kaban, J. Bednarčik, A. Pradel, Short range order and stability of amorphous $Ge_xTe_{100-x}$ alloys (12 <= x <= 44.6), J. Phys.: Condens. Matter 25 (2013) 195401. doi:10.1088/0953-8984/25/19/195401.

[25] G. Wang, Q. Nie, M. Barj, X. Wang, S. Dai, X. Shen, et al., Compositional dependence of the optical properties of novel Ge–Ga–Te–CsI far infrared transmitting chalcohalide glasses system, J. Phys. Chem. Solids 72 (2011) 5–9. doi:10.1016/j.jpcs.2010.10.018.

[26] R. Golovchak, L. Calvez, B. Bureau, H. Jain, Structural evolution of Ga-Ge-Te glasses by combined EXAFS and XPS analysis, J. Chem. Phys. 139 (2013) 54508.

[27] V.S. Shiryaev, M.F. Churbanov, Trends and prospects for development of chalcogenide fibers for mid-infrared transmission, J. Non-Cryst. Solids 377 (2013) 225–230. doi:10.1016/j.jnoncrysol.2012.12.048.

[28] Y. He, X. Wang, Q. Nie, Y. Xu, T. Xu, S. Dai, Glass formation and optical properties of Ge–Te–Ga–CuI far-IR transmitting chalcogenide glasses, Infrared Phys. Technol. 60 (2013) 129–133. doi:10.1016/j.infrared.2013.04.008.

[29] M. Hubert, E. Petracovschi, X.-H. Zhang, L. Calvez, Synthesis of Germanium-Gallium-Tellurium (Ge-Ga-Te) Ceramics by Ball-Milling and Sintering, J. Am. Ceram. Soc. 96 (2013) 1444–1449. doi:10.1111/jace.12299.

[30] https://www.webelements.com/

[31] Y. Kameda, O. Uemura, T. Usuki, Time-of-Flight Neutron Diffraction Study of Amorphous and Liquid Ge-Te Alloys, Mater. Trans. 37 (1996) 1655.

[32] K. Hirota, K. Nagino, G. Ohbayashi, Local structure of amorphous GeTe and PdGeSbTe alloy for phase change optical recording, J. Appl. Phys. 82 (1997) 65–70. doi:10.1063/1.365850.

[33] M. Sakurai, F. Kakinuma, E. Matsubara, K. Suzuki, Partial structure analysis of amorphous $Ge_{15}Te_{80}M_5$ (M=Cu, Ag and In), J. Non-Cryst. Solids 312–314 (2002) 585–588. doi:10.1016/S0022-3093(02)01789-1.

[34] S. Sen, S. Joshi, B.G. Aitken, S. Khalid, Atomic structure and chemical order in binary Ge–Te and As–Te glasses: A Te K-edge X-ray absorption fine structure spectroscopic study, J. Non-Cryst. Solids 354 (2008) 4620–4625. doi:10.1016/j.jnoncrysol.2008.05.048.

[35] Y.G. Choi, S.Y. Shin, R. Golovchak, B. Cheong, H. Jain, Comparative study of atomic arrangements in equiatomic GeSe and GeTe films before and after crystallization, J. Alloys Compd. 686 (2016) 273–280. doi:10.1016/j.jallcom.2016.06.021.

[36] P. Noé, C. Sabbione, N. Castellani, G. Veux, G. Navarro, V. Sousa, et al., Structural change with the resistance drift phenomenon in amorphous GeTe phase change materials' thin films, J. Phys. D: Appl. Phys. 49 (2016) 035305. doi:10.1088/0022-3727/49/3/035305.

[37] I. Kaban, P. Jóvári, W. Hoyer, E. Welter, Determination of partial pair distribution functions in amorphous $Ge_{15}Te_{85}$ by simultaneous RMC simulation of diffraction and EXAFS data, J. Non-Cryst. Solids 353 (2007) 2474–2478. doi:10.1016/j.jnoncrysol.2006.09.072.

[38] A.A. Piarristeguy, M. Micoulaut, R. Escalier, P. Jóvári, I. Kaban, J. van Eijk, et al., Structural singularities in $Ge_xTe_{100-x}$ films, J. Chem. Phys. 143 (2015) 074502. doi:10.1063/1.4928504.

[39] J.R. Stellhorn, S. Hosokawa, W.-C. Pilgrim, N. Blanc, N. Boudet, H. Tajiri, et al., Short- and intermediate-range order in amorphous GeTe, Phys. Status Solidi B 253 (2016) 1038–1045. doi:10.1002/pssb.201552559.

[40] A. V Kolobov, P. Fons, A.I. Frenkel, A.L. Ankudinov, J. Tominaga, T. Uruga, Understanding the phase-change mechanism of rewritable optical media, Nat. Mater. 3 (2004) 703–708. doi:10.1038/nmat1215.

[41] P. Jóvári, I. Kaban, J. Steiner, B. Beuneu, A. Schöps, A. Webb, "Wrong bonds" in sputtered amorphous $Ge_2Sb_2Te_5$, J. Phys.: Condens. Matter 19 (2007) 335212. doi:10.1088/0953-8984/19/33/335212.

[42] P. Jóvári, I. Kaban, J. Steiner, B. Beuneu, A. Schöps, M. Webb, Local order in amorphous $Ge_2Sb_2Te_5$ and $GeSb_2Te_4$, Phys. Rev. B 77 (2008) 035202. doi:10.1103/PhysRevB.77.035202.

[43] J. Akola, R.O. Jones, Structural phase transitions on the nanoscale: The crucial pattern in the phase-change materials $Ge_2Sb_2Te_5$ and GeTe, Phys. Rev. B 76 (2007) 235201. doi:10.1103/PhysRevB.76.235201.

[44] J. Hegedüs, S.R. Elliott, Microscopic origin of the fast crystallization ability of Ge–Sb–Te phase-change memory materials, Nat. Mater. 7 (2008) 399–405. doi:10.1038/nmat2157.

[45] J. Akola, R.O. Jones, Structure of amorphous $Ge_8Sb_2Te_{11}$ : $GeTe$-$Sb_2Te_3$ alloys and optical storage, Phys. Rev. B 79 (2009) 134118. doi:10.1103/PhysRevB.79.134118.

[46] J. Kalikka, J. Akola, R.O. Jones, S. Kohara, T. Usuki, Amorphous $Ge_{15}Te_{85}$ : density functional, high-energy x-ray and neutron diffraction study, J. Phys.: Condens. Matter 24 (2012) 015802. doi:10.1088/0953-8984/24/1/015802.

[47] K. Gunasekera, P. Boolchand, M. Micoulaut, Effect of mixed Ge/Si cross-linking on the physical properties of amorphous Ge-Si-Te networks, J. Appl. Phys. 115 (2014) 164905. doi:10.1063/1.4871780.

[48] M. Micoulaut, K. Gunasekera, S. Ravindren, P. Boolchand, Quantitative measure of tetrahedral-sp3 geometries in amorphous phase-change alloys, Phys. Rev. B 90 (2014) 094207. doi:10.1103/PhysRevB.90.094207.

[49] M. Micoulaut, A. Piarristeguy, H. Flores-Ruiz, A. Pradel, Towards accurate models for amorphous GeTe: Crucial effect of dispersive van der Waals corrections on the structural properties involved in the phase-change mechanism, Phys. Rev. B 96 (2017) 184204. doi:10.1103/PhysRevB.96.184204.

[50] A. Bouzid, G. Ori, M. Boero, E. Lampin, C. Massobrio, Atomic-scale structure of the glassy $Ge_2Sb_2Te_5$ phase change material: A quantitative assessment via first-principles mole, Phys. Rev. B 96 (2017) 224204. doi:10.1103/PhysRevB.96.224204.

[51] J. Akola, R.O. Jones, Density functional study of amorphous, liquid and crystalline $Ge_2Sb_2Te_5$: homopolar bonds and/or AB alternation?, J. Phys.: Condens. Matter 20 (2008) 465103. doi:10.1088/0953-8984/20/46/465103.

[52] A.M. Loireau-Lozac'h, F. Keller-Besrest, S. Bénazeth, Short and medium range order in Ga-Ge-S glasses: An X-ray absorption spectroscopy study at room and low temperatures, J. Solid State Chem. 123 (1996) 60–67. doi:10.1006/jssc.1996.0152.

[53] J.H. Song, Y.G. Choi, J. Heo, Ge and Ga K-edge EXAFS analyses on the structure of Ge-Ga-S-CsBr glasses, J. Non-Cryst. Solids 352 (2006) 423–428. doi:10.1016/j.jnoncrysol.2006.01.013.

[54] A.S. Tverjanovich, E. V Tereshchenko, Structural investigation of glasses in the $x(0.16GaCh_2 \cdot 0.84GeCh_2) \cdot (1 − x)(SbCh_{1.5})$ (Ch = S, Se) system, Glass Phys. Chem. 35 (2009) 475–478. doi:10.1134/S1087659609050046.

[55] Y. Ledemi, S.H. Messaddeq, I. Skhripachev, S.J.L. Ribeiro, Y. Messaddeq, Influence of Ga incorporation on photoinduced phenomena in Ge-S based glasses, J. Non-Cryst. Solids 355 (2009) 1884–1889. doi:10.1016/j.jnoncrysol.2009.04.046.

[56] C. Lin, L. Calvez, H. Tao, M. Allix, A. Moréac, X.H. Zhang, et al., Evidence of network demixing in $GeS_2$-$Ga_2S_3$ chalcogenide glasses: A phase transformation study, J. Solid State Chem. 184 (2011) 584–588. doi:10.1016/j.jssc.2011.01.018.

[57] P. Masselin, D. Le Coq, A. Cuisset, E. Bychkov, Spatially resolved Raman analysis of laser induced refractive index variation in chalcogenide glass, Opt. Mater. Express 2 (2012) 1768. doi:10.1364/OME.2.001768.

[58] I. Pethes, V. Nazabal, R. Chahal, B. Bureau, I. Kaban, S. Belin, et al., Local motifs in $GeS_2$–$Ga_2S_3$ glasses, J. Alloys Compd. 673 (2016) 149–157. doi:http://dx.doi.org/10.1016/j.jallcom.2016.02.193.

[59] E. Zhu, X. Zhao, J. Wang, C. Lin, Compositional dependence of physical and structural properties in $(Ge_{1-x}Ga_x)S_2$ chalcogenide glasses, J. Non-Cryst. Solids 489 (2018) 45–49. doi:10.1016/j.jnoncrysol.2018.03.030.

[60] P. Němec, B. Frumarová, M. Frumar, Structure and properties of the pure and Pr3+-doped $Ge_{25}Ga_5Se_{70}$ and $Ge_{30}Ga_5Se_{65}$ glasses, J. Non-Cryst. Solids 270 (2000) 137–146. doi:10.1016/S0022-3093(00)00066-1.

[61] A.W. Mao, B.G. Aitken, R.E. Youngman, D.C. Kaseman, S. Sen, Structure of Glasses in the Pseudobinary System $Ga_2Se_3$-$GeSe_2$: Violation of Chemical Order and 8-N Coordination Rule, J. Phys. Chem. B 117 (2013) 16594–16601. doi:10.1021/jp410017k.

[62] R. Golovchak, L. Calvez, E. Petracovschi, B. Bureau, D. Savytskii, H. Jain, Incorporation of Ga into the structure of Ge–Se glasses, Mater. Chem. Phys. 138 (2013) 909–916. doi:10.1016/j.matchemphys.2012.12.084.

[63] Y. Ledemi, B. Bureau, G. Le Caër, L. Calvez, C. Roiland, G. Tricot, et al., $^{71}$Ga NMR in chalcogenide and chalco-halide glasses, J. Non-Cryst. Solids 383 (2014) 216–221. doi:10.1016/j.jnoncrysol.2013.03.038.

[64] I. Pethes, R. Chahal, V. Nazabal, C. Prestipino, A. Trapananti, C. Pantalei, et al., Short range order in Ge-Ga-Se glasses, J. Alloys Compd. 651 (2015) 578–584. doi:10.1016/j.jallcom.2015.08.039.

[65] B.D. Klee, J.R. Stellhorn, M. Krbal, N. Boudet, G.A. Chahine, N. Blanc, et al., Distinguishing the local environments of Ga and Ge in amorphous $(Ga_2Se_3)_{0.25}(GeSe_2)_{0.75}$ by anomalous X-ray scattering, Chalcogenide Lett. 15 (2018) 1–6.

[66] E. Petracovschi, L. Calvez, L. Cormier, D. Le Coq, J. Du, Short and medium range structures of $80GeSe_2$ –$20Ga_2Se_3$ chalcogenide glasses, J. Phys.: Condens. Matter 30 (2018) 185403. doi:10.1088/1361-648X/aaaf36.

**Tables**

**Table 1** Estimated densities, number densities, fitted experimental data sets.

|  | $\rho$ [g/cm$^3$] | Number density [Å$^{-3}$] | Experimental data sets |
|---|---|---|---|
| Ge$_{7.5}$Ga$_{7.5}$Te$_{85}$ | 5.6 | 0.0283 | ND, XRD, Ge, Ga EXAFS |
| Ge$_{10}$Ga$_{10}$Te$_{80}$ | 5.57 | 0.0288 | ND, XRD |
| Ge$_{12.5}$Ga$_{12.5}$Te$_{75}$ | 5.53 | 0.0293 | ND, XRD, Ge, Ga EXAFS |
| Ge$_{14.3}$Ga$_{14.3}$Te$_{71.4}$ | 5.7 | 0.0308 | ND, XRD, Ge, Ga EXAFS |

**Table 2** Minimum interatomic distances (in Å) used in the reverse Monte Carlo simulation runs.

|  | Ge-Ge | Ge-Ga | Ge-Te | Ga-Ga | Ga-Te | Te-Te |
|---|---|---|---|---|---|---|
| Bond allowed | 2.35 | 2.35 | 2.4 | 2.35 | 2.4 | 2.5 |
| Bond forbidden | 3.45 | 3.45 |  | 3.45 |  |  |

**Table 3** Nearest neighbor distances (in Å). The uncertainty of distances is about ± 0.02 Å if EXAFS data sets are also fitted, but it can be higher (about 0.05 Å) in case of the Ge$_{10}$Ga$_{10}$Te$_{80}$, where only the two diffraction measurements were available.

|  | Ge-Te | Ga-Te | Te-Te |
|---|---|---|---|
| Ge$_{7.5}$Ga$_{7.5}$Te$_{85}$ | 2.60 | 2.62 | 2.78 |
| Ge$_{10}$Ga$_{10}$Te$_{80}$ | 2.64 | 2.64 | 2.74 |
| Ge$_{12.5}$Ga$_{12.5}$Te$_{75}$ | 2.60 | 2.63 | 2.77 |
| Ge$_{14.3}$Ga$_{14.3}$Te$_{71.4}$ | 2.61 | 2.62 | 2.77 |

**Table 4** Coordination numbers of the investigated glasses obtained by simulations for the final model, in which only the Ge-Te, Ga-Te and Te-Te bonds were allowed and the average coordination numbers were not constrained.

| Pair (upper limit) | Ge$_{7.5}$Ga$_{7.5}$Te$_{85}$ | Ge$_{10}$Ga$_{10}$Te$_{80}$ | Ge$_{12.5}$Ga$_{12.5}$Te$_{75}$ | Ge$_{14.3}$Ga$_{14.3}$Te$_{71.4}$ |
|---|---|---|---|---|
| $N_{Ge-Te}$ (3.1 Å) | 4.0 (-0.4+0.6) | 4.1 (-0.4+0.7) | 3.9 (-0.4+0.5) | 3.95 (±0.3) |
| $N_{Te-Ge}$ (3.1 Å) | 0.35 (-0.03+0.06) | 0.51 (-0.05+0.1) | 0.65 (±0.08) | 0.79 (±0.06) |
| $N_{Ga-Te}$ (3.1 Å) | 3.6 (-0.6+0.4) | 3.8 (-0.2+1.0) | 3.8 (-0.4+0.6) | 3.9 (-0.4+0.5) |
| $N_{Te-Ga}$ (3.1 Å) | 0.32(-0.06+0.03) | 0.48 (-0.03+0.1) | 0.63 (±0.08) | 0.78(±0.08) |
| $N_{Te-Te}$ (3.0 Å) | 1.4 (±0.06) | 1.1 (-0.1+0.05) | 0.91 (-0.2+0.1) | 0.78 (-0.15+0.1) |
| $N_{Te}$ | 2.07 (±0.07) | 2.1 (-0.1+0.05) | 2.19 (-0.15+0.1) | 2.35 (±0.1) |

**Table 5** Percentage of Ge and Ga atoms participating in corner or/and edge sharing units. Uncertainties were determined from 10 simulation runs started from different initial configurations.

| | $Ge_{7.5}Ga_{7.5}Te_{85}$ | $Ge_{10}Ga_{10}Te_{80}$ | $Ge_{12.5}Ga_{12.5}Te_{75}$ | $Ge_{14.3}Ga_{14.3}Te_{71.4}$ |
|---|---|---|---|---|
| Neither CS nor ES Ge | 16.5 (±1) | 6.8 | 1.5 | 0.3 (±0.1) |
| Neither CS nor ES Ga | 16.5 (±2) | 5.6 | 1.1 | 0.25 (±0.15) |
| Only CS Ge | 71.5 (±1.5) | 75 | 71 | 65 (±1.5) |
| Only CS Ga | 71.5 (±2) | 75 | 71 | 67 (±1) |
| Only ES Ge | 3.2 (±0.5) | 2.9 | 1.6 | 0.8 (±0.2) |
| Only ES Ga | 3.2 (±0.5) | 2.3 | 1.5 | 0.5 (±0.2) |
| CS and ES Ge | 9 (±1) | 15 | 26 | 34 (±2) |
| CS and ES Ga | 9 (±1) | 17 | 26 | 32 (±1) |

**Table 6** Position ($q_{max}$) and amplitude of the first sharp diffraction peak (see text for definition).

| | $q_{max}$ [Å$^{-1}$] | $q_{min}$ [Å$^{-1}$] | $S(q_{max})$-$S(q_{min})$ |
|---|---|---|---|
| $Ge_{7.5}Ga_{7.5}Te_{85}$ | 1.01 | 1.33 | -0.003 |
| $Ge_{10}Ga_{10}Te_{80}$ | 1.01 | 1.33 | 0.13 |
| $Ge_{12.5}Ga_{12.5}Te_{75}$ | 1.01 | 1.33 | 0.198 |
| $Ge_{14.3}Ga_{14.3}Te_{71.4}$ | 0.93 | 1.33 | 0.256 |

**Figures**

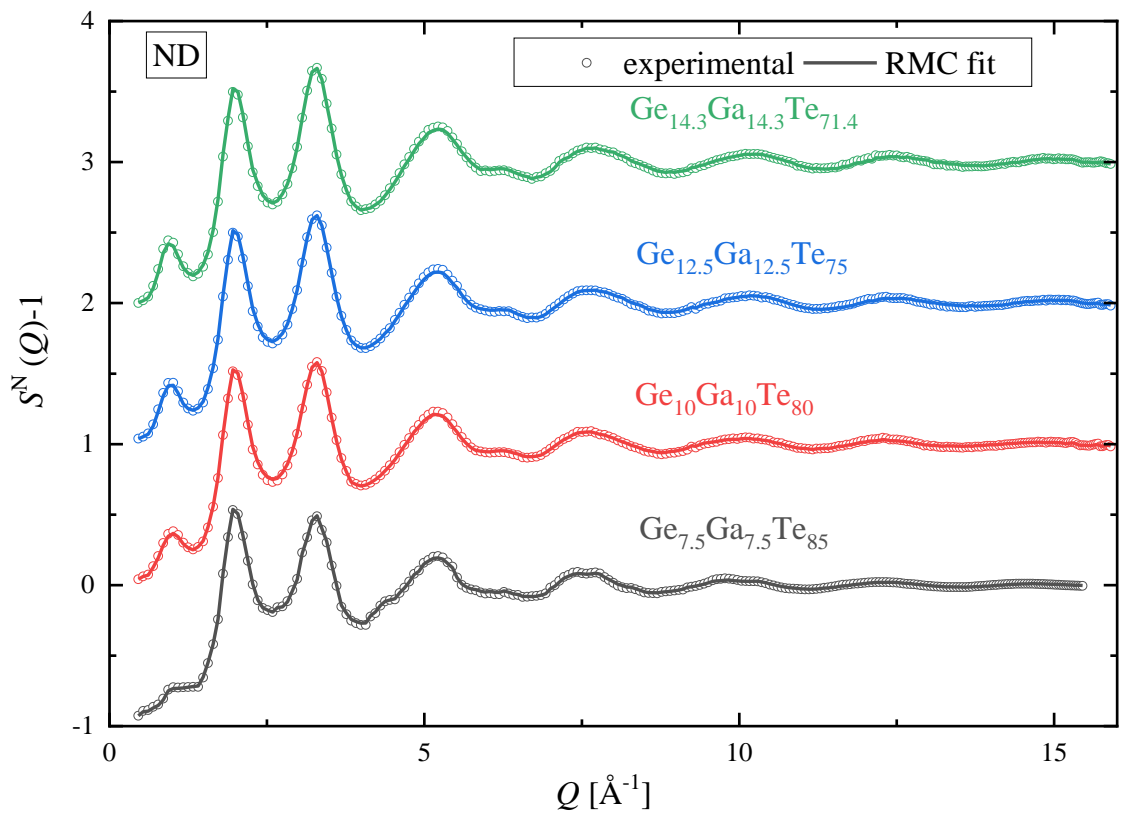

**Figure 1**. ND structure factors (symbols) and fits (lines) of the $Ge_xGa_xTe_{100-2x}$ glasses. (The curves are shifted vertically for clarity.)

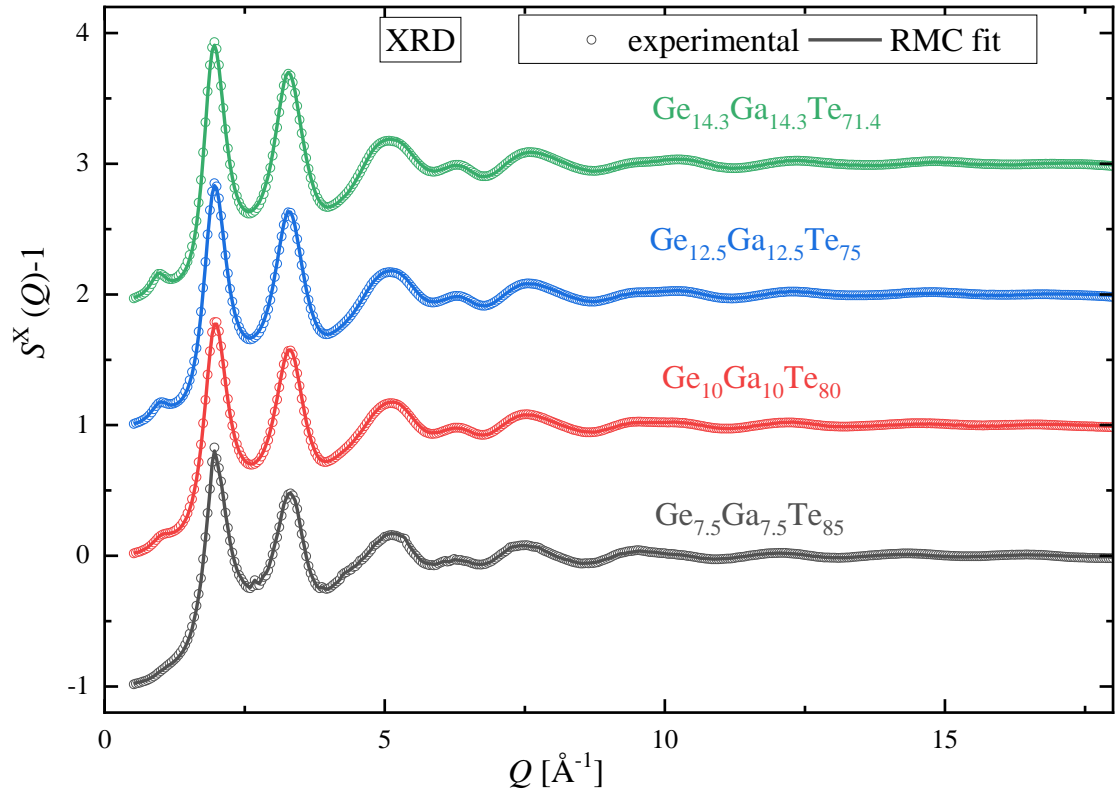

**Figure 2.** XRD structure factors (symbols) and fits (lines) of the $Ge_xGa_xTe_{100-2x}$ glasses. (The curves are shifted vertically for clarity.)

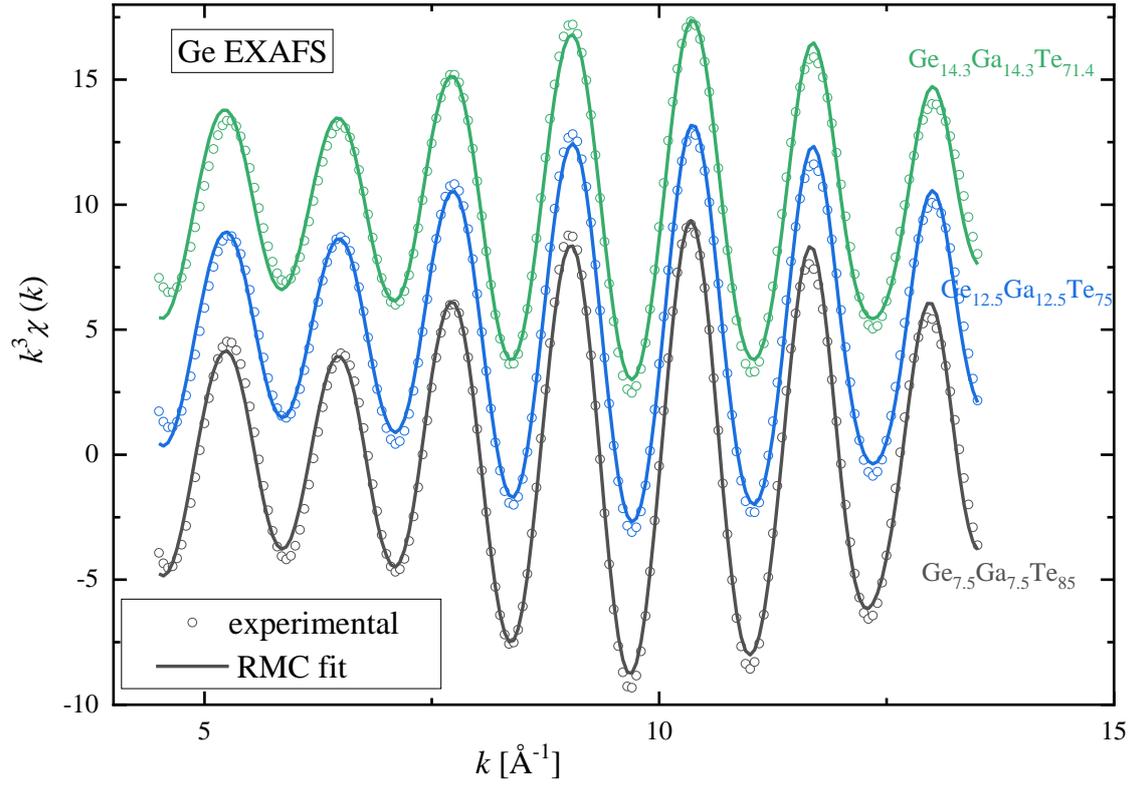

**Figure 3**. $k^3$-weighted, filtered EXAFS spectra at Ge K-edge (symbols) and fits (lines) of the Ge$_x$Ga$_x$Te$_{100-2x}$ glasses. (The curves are shifted vertically for clarity.)

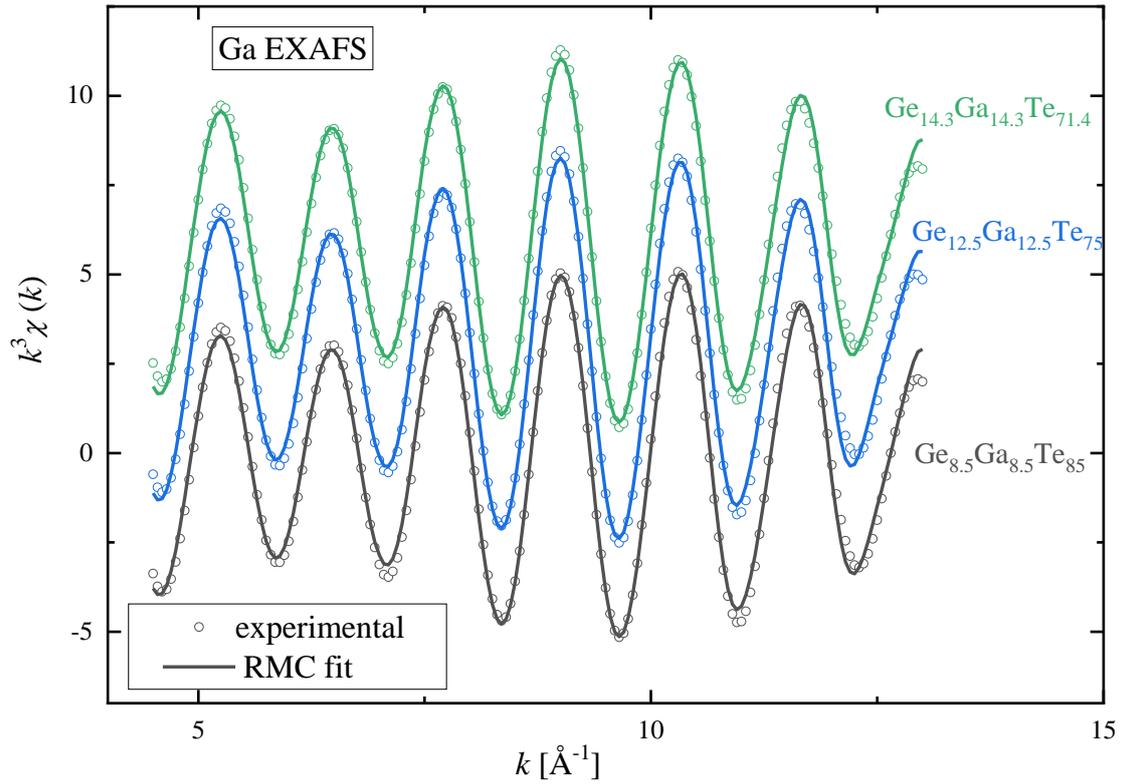

**Figure 4**. $k^3$-weighted, filtered EXAFS spectra at Ga K-edge (symbols) and fits (lines) of the Ge$_x$Ga$_x$Te$_{100-2x}$ glasses. (The curves are shifted vertically for clarity.)

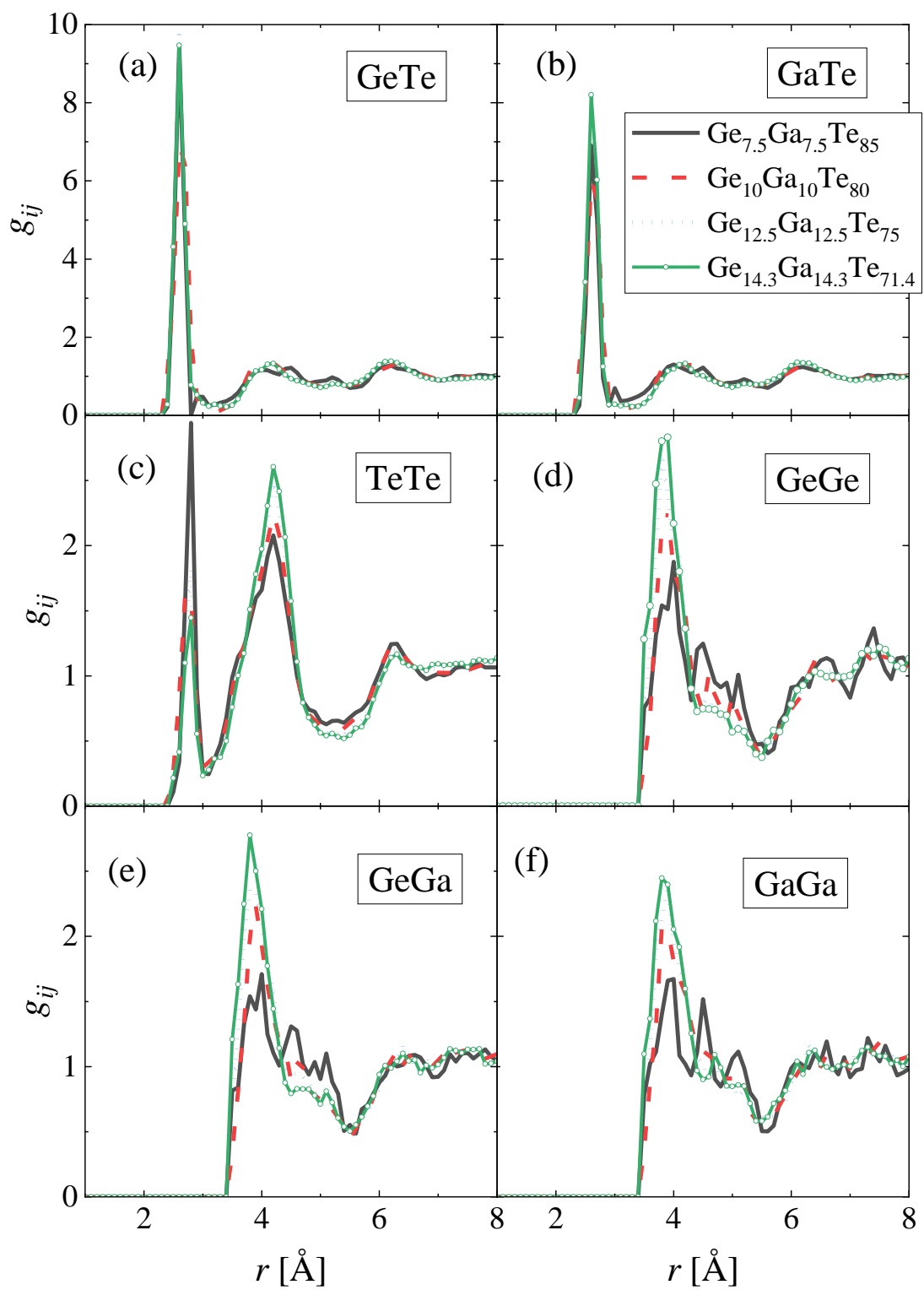

**Figure 5**. Partial pair correlation functions of the Ge$_x$Ga$_x$Te$_{100-2x}$ glasses.

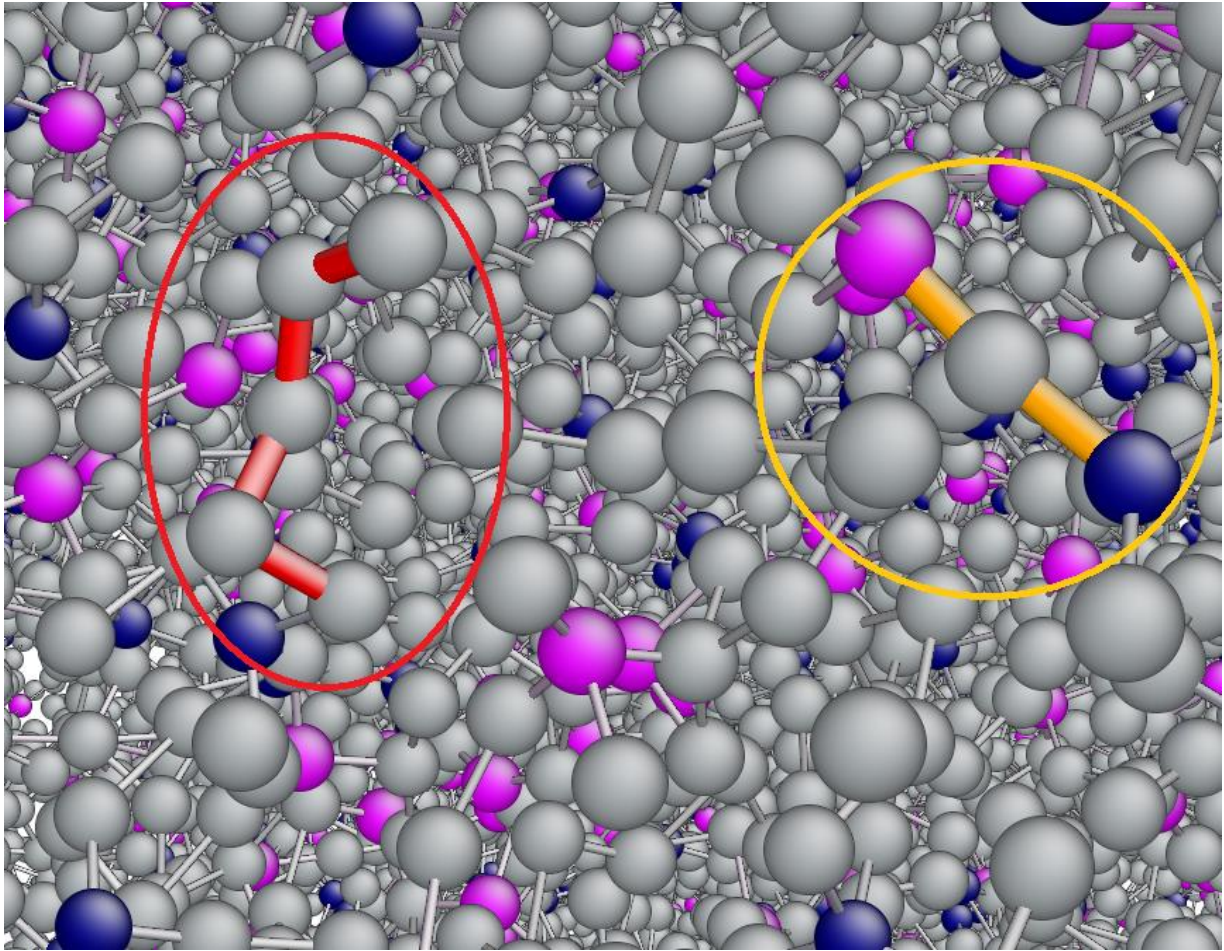

**Figure 6**. A snapshot about a part of the configuration of the Ge$_{7.5}$Ga$_{7.5}$Te$_{85}$ glass obtained by RMC simulation. The Ge, Ga and Te atoms are represented by magenta, blue and grey balls, respectively. Two corner sharing tetrahedra are marked with orange, a short chain of Te atoms is highlighted by red.

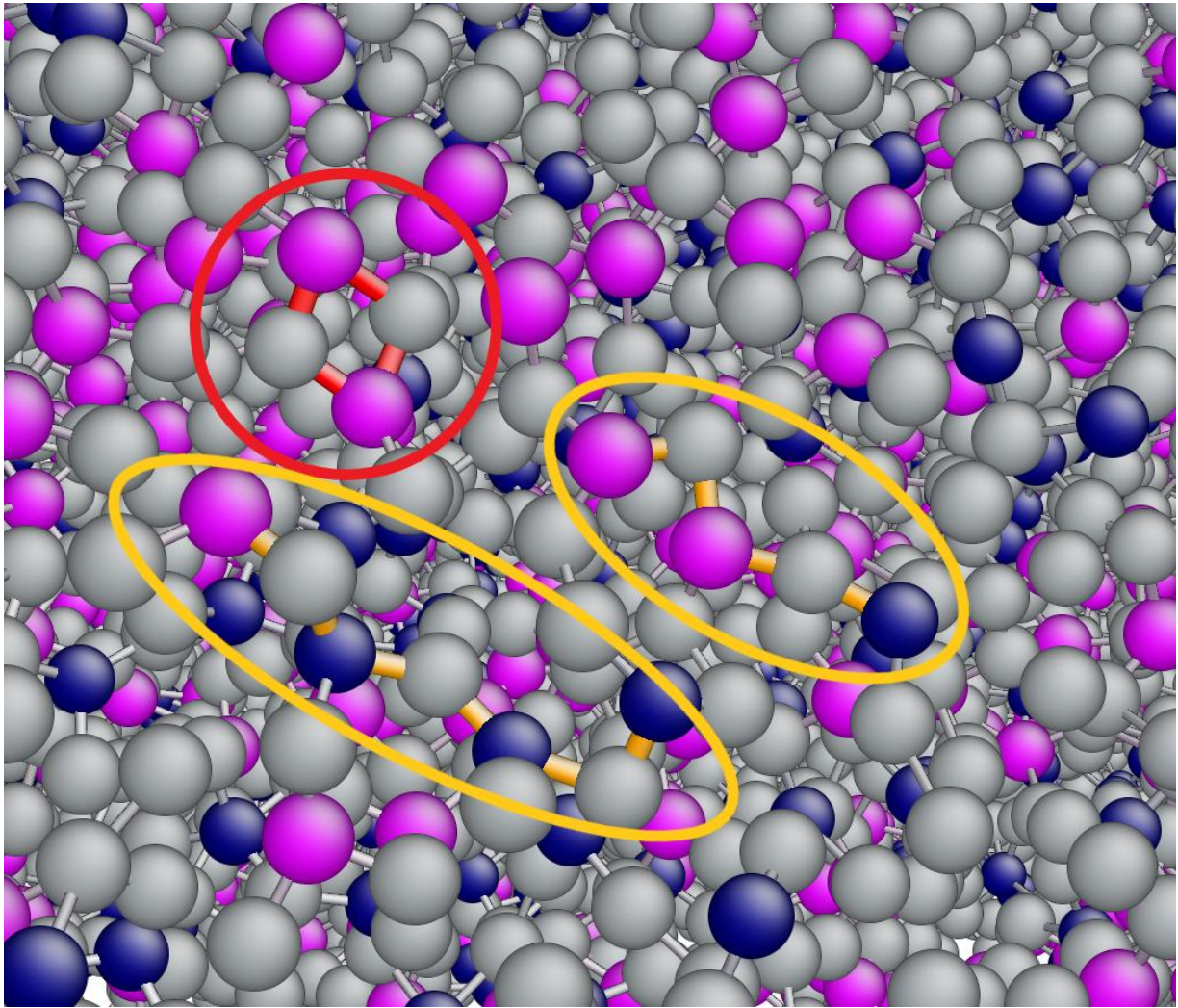

**Figure 7**. A snapshot about a part of the configuration of the Ge$_{14.3}$Ga$_{14.3}$Te$_{71.4}$ glass obtained by RMC simulation. The Ge, Ga and Te atoms are represented by magenta, blue and grey balls, respectively. Two edge-sharing GaTe$_4$ tetrahedra are marked by red; chains of corner-sharing MTe$_4$ units are highlighted with orange.

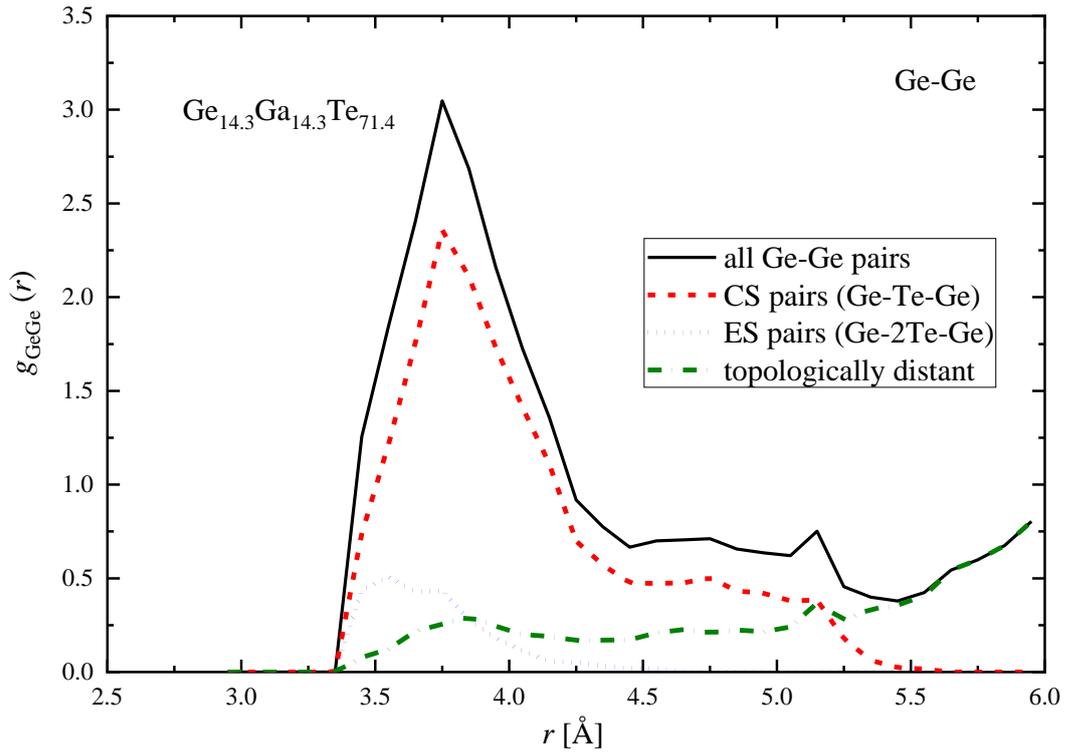

**Figure 8**. Decomposition of $g_{GeGe}(r)$ of Ge$_{14.3}$Ga$_{14.3}$Te$_{71.4}$ glass to contributions from corner (CS) and edge (ES) sharing tetrahedra and topologically distant Ge-Ge pairs.

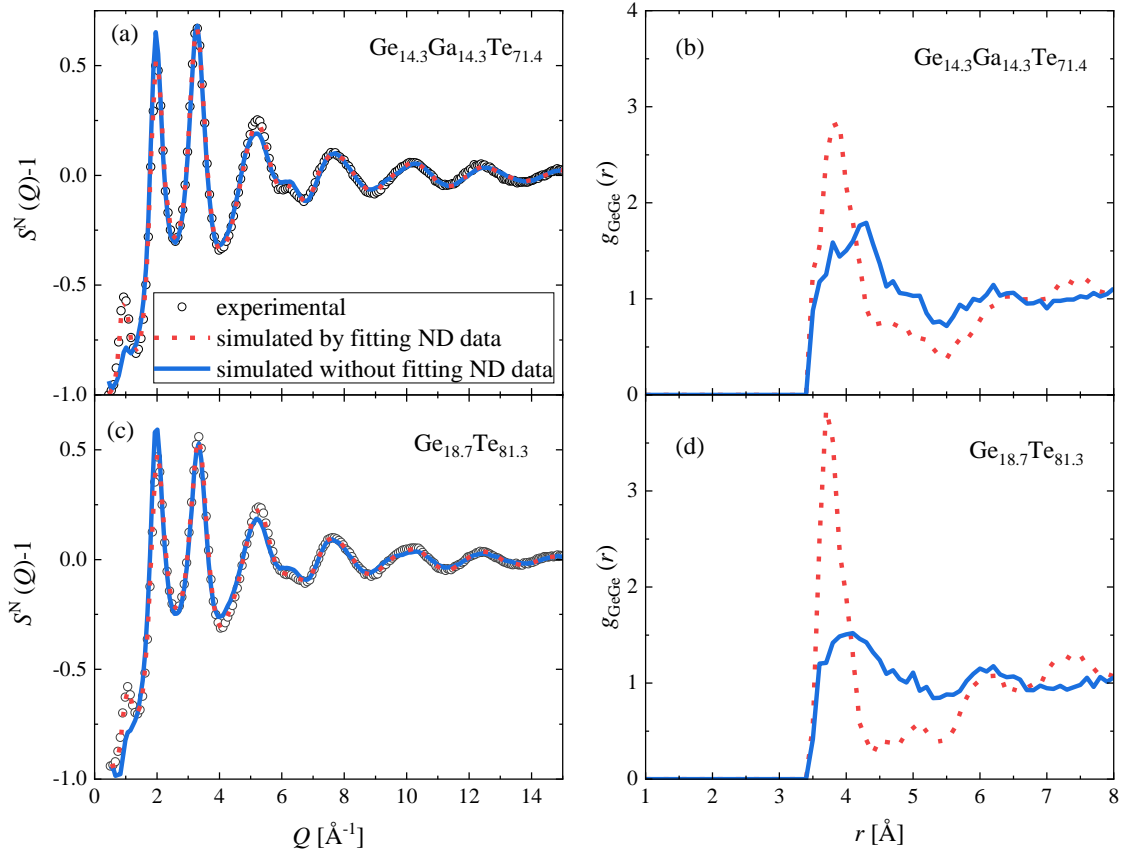

**Figure 9**. Comparison of the neutron weighted structure factors (a and c) and the first peak of the Ge-Ge partial pair correlation functions (b and d) of $Ge_{14.3}Ga_{14.3}Te_{71.4}$ and $Ge_{18.7}Te_{81.3}$ glasses obtained from RMC simulations by (red lines) fitting the experimental neutron diffraction data and (blue line) without fitting ND data. Experimental data (symbols) are also shown for reference.